# High-power, hybrid Er:fiber/Tm:fiber frequency comb source in the 2 µm wavelength region

Florian Adler[1,2] and Scott A. Diddams[1,*]

[1]*National Institute of Standards and Technology, Time and Frequency Division,*
*Mailstop 847.00, 325 Broadway, Boulder, Colorado 80305, USA*
[2]*University of Colorado, Boulder, Colorado 80309, USA*
*\*Corresponding author: sdiddams@boulder.nist.gov*

We present a 2-µm frequency comb based on a reliable mode-locked Er:fiber laser with 100 MHz repetition rate. After shifting the spectrum of the amplified Er:fiber comb to longer wavelengths, a single-clad Tm/Ho:fiber is used as a self-pumped pre-amplifier to generate a coherent and broadband spectrum centered at 1.93 µm. Subsequently, a cladding-pumped Tm:fiber amplifier boosts the system to a maximum output power of 4.8 W at 1.96 µm. After compression in a compact grating compressor, our amplified Er:fiber/Tm:fiber hybrid system delivers as much as 2.9 W with a pulse duration of 141 fs. The system's comb properties are examined via heterodyne measurement.

Optical frequency combs have made a significant impact in recent years as sources for broadband and high-resolution molecular spectroscopy [1–6]. For future applications in atmospheric science, trace gas detection, and homeland security, the mid-infrared spectral region seems to be particularly promising, pushing the development of frequency comb sources with wavelengths longer than 2.5 µm. Employing optical parametric frequency conversion of near-IR systems such as Yb:fiber ($\lambda$ = 1.06 µm) or Er:fiber ($\lambda$ = 1.55 µm) lasers, several solutions targeting the 2.5-to-5-µm wavelength range have been successfully implemented [7–12]. First spectroscopy experiments were able to show the outstanding versatility and sensitivity of these novel mid-infrared systems for important molecular absorption bands such as the C-H-stretch region and the fundamental vibration bands of $CO_2$ and $N_2O$ [13].

To further increase the applicability of frequency comb spectroscopy, it is desirable to extend the wavelength coverage farther into the infrared (>5 µm) to obtain higher sensitivity for certain molecules (e.g., $H_2O$, NO, and $O_3$) and to target more complex species such as halocarbons or explosives. Whereas current mid-IR comb sources rely mostly on periodically poled $LiNbO_3$ (PPLN) crystals for frequency conversion, long-wave mid-IR systems will have to employ alternate crystals such as orientation-patterned GaAs (OP-GaAs), silver gallium selenide (AGSE), or zinc germanium phosphide (ZGP) to overcome the ~5 µm transparency limit of PPLN. In turn, these crystals rule out the use of pump sources at 1 µm or 1.5 µm, owing to multiphoton absorption at these short wavelengths. Therefore, the development of suitable pump sources at longer wavelengths is a key step toward extending the coverage of mid-infrared frequency combs. Current work focuses on the development of frequency combs based on free-space Ho:YLF ($\lambda$ = 2.06 µm) [14] or Cr:ZnSe lasers ($\lambda$ = 2.45 µm) [10], as well as femtosecond Tm:fiber lasers ($\lambda$ = 1.95 µm) [15]; however, all of these systems are still experimental and not commercially available.

In this Letter, we present an alternate approach for the generation of a 2 µm frequency comb based on a reliable and commercially available Er:fiber oscillator. We generate a high-quality frequency comb around 2 µm by shifting the spectrum of the femtosecond Er:fiber laser further into the infrared via the addition of a nonlinear fiber and a Tm/Ho:fiber. Subsequently, the laser light is amplified in a cladding-pumped, polarization-maintaining (PM) Tm:fiber amplifier to generate a high-power frequency comb centered at 1.96 µm, which represents a suitable seed source for optical parametric processes targeting the 3 µm to 12 µm wavelength range.

Figure 1(a) shows a schematic of the seed laser system. A commercial femtosecond Er:fiber oscillator with a repetition rate of 100 MHz (Menlo Systems C-Comb) generates the original frequency comb centered around 1.575 µm. After a single-mode Er:fiber amplifier stage pumped with three 750 mW, 980 nm pump diodes, the output pulse energy is 5.5 nJ. A piece of SMF-28 for pulse compression is directly spliced on, followed by 10 cm of highly nonlinear fiber (HNF) with a core diameter of ~4 µm and 30 cm of single-clad Tm/Ho:fiber (9 µm core diameter). The HNF produces a supercontinuum exhibiting a spectral peak around $\lambda \approx 2.0$ µm that contains approximately 40 mW of total power. The short piece of HNF is not able to concentrate a majority of the power in the desired wavelength region; however, the rapid spectral broadening, which relies mostly on four-wave mixing, ensures that the shifted spectrum is still entirely coherent [16,17], as opposed to the more efficient Raman self-frequency shift [18], which may compromise the comb's coherence [11]. In order to generate a frequency comb with sufficient quality for direct spectroscopy applications, preserving the coherence is highly important. Although in our approach most of the spectral power remains near the original wavelength between 1.5 µm and 1.6 µm, this light can be utilized to pump Tm-doped fibers. Therefore, the addition of a piece of Tm/Ho-doped single-clad fiber acts as a self-pumped pre-amplifier [19], boosting the power at the desired output wavelength without requiring an additional pump source. After coupling the light into free-space, the measured power is 150 mW in a well-isolated broadband spectrum centered around 1.93 µm. To ensure that this output is suitable to seed a high-power amplifier, we investigate its time-domain characteristics by compressing the pulses with a silicon prism pair, which provides a tunable amount of positive dispersion. Although the compressor arrangement was not able to compensate high dispersion orders over the ~200 nm total bandwidth of the spectrum, we are able to obtain a pulse duration of only 60 fs. The temporal and spectral pulse characteristics were measured via SHG-FROG [20] and are shown in Figures 1(b) and 1(c), respectively. We can therefore conclude that the pulse characteristics of our hybrid seed laser are comparable to those of recently developed mode-locked Tm:fiber oscillators [15], making it a viable alternative as a 2 μm frequency comb source.

As a next step, the output power of our hybrid system is increased by employing a double-clad Tm:fiber amplifier stage. A schematic of the amplifier is shown in Figure 2(a). The pulses from the seed laser are first coupled into 3.0 m of undoped fiber for pre-stretching ($\beta_2 = -100$ ps$^2$/km); afterwards they enter 2.8 m of double-clad Tm-doped fiber that is immersed in inert cooling fluid. The active fiber has 10 μm core and 130 μm cladding diameter, respectively, and is counter-pumped by a 16 W, 793 nm multimode laser diode through a pump-signal combiner (PSC). The amplifier is entirely fiber-coupled and polarization-maintaining for optimum long-term stability. The active fiber is cooled to −8°C, yielding a maximum output power of 4.8 W, which is coupled out into free-space via an integrated fiber isolator and collimator (ICM). A compact, folded grating compressor combining diffractive and focusing optics [21,22] provides positive dispersion to compress the pulses from the high-power amplifier. We achieve a throughput of >60% (limited by the reflectivity of the gold coatings on the mirrors and the grating), resulting in a maximum compressed output power of 2.9 W centered at 1.96 μm. The pulse characteristics are again measured via SHG-FROG, and we obtain a pulse duration of 141 fs, which is limited by uncompensated higher orders of dispersion [see Figures 2(b) and (c)].

We also analyze the relative intensity noise (RIN) of the amplified system. At maximum power, the integrated rms amplitude noise from 1 MHz to 1 Hz amounts to $7 \times 10^{-3}$. Most of this noise is located at higher frequencies and originates from the nonlinear light generation in the HNF. In addition, the Tm:fiber amplifier itself may cause low-frequency noise or drift due to temperature instabilities. There may also be a drift in output power caused by polarization changes in the non-PM Er:fiber amplifier. Therefore, we measure the power stability of the system by recording the output with a thermal power meter and an oscilloscope over 500 s (maximum recording period of the oscilloscope). This measurement reveals relative rms fluctuations of $1.8 \times 10^{-3}$; no discernible drift of the output power is present over the recorded timespan.

Due to the nonlinear nature of the pulse generation and the high intensity inside the Tm:fiber amplifier, there is some concern that the frequency comb coherence could be compromised. To use the laser as a future pump source for mid-infrared combs, this aspect is extremely critical. Therefore, we examine the frequency noise characteristics of the system by generating a heterodyne beat between a frequency-stabilized 976 nm cw laser and the frequency-doubled output of the double-clad PM-Tm:fiber amplifier. The noise data are recorded by feeding the beat note signal into a frequency-to-voltage converter and analyzing its output with a vector signal analyzer. The resulting frequency noise spectral density $S_\nu(f)$ is shown in Figure 3. The inset shows the beat note on an rf spectrum analyzer. When integrated from 1 MHz to 10 Hz, the total frequency noise amounts to 107 kHz. By assuming that the noise from the cw laser is negligible, and taking into account that we measure the second harmonic of the Tm:fiber system, we find an upper limit of 54 kHz for the comb's free-running integrated frequency noise. This result leads to the conclusion that the high nonlinearities in the system add some excess noise; however, the frequency comb structure is essentially intact and fully coherent, as required for the system's intended use.

In summary, we demonstrate that converting an Er:fiber frequency comb to a 2 μm source provides an excellent and uncomplicated solution for the generation of a coherent frequency comb in this wavelength region that is competitive to recently developed systems based solely on Tm:fiber technology [15]. Furthermore, this design offers the proven reliability and availability of the well-established femtosecond Er:fiber lasers. After amplification in polarization-maintaining, double-clad Tm:fiber, our compressed pulses exhibit a peak power of more than 200 kW (29 nJ pulse energy); the system is therefore an excellent pump source for infrared nonlinear fibers and future long-wave mid-IR parametric frequency combs based on OP-GaAs, AGSE, or ZGP crystals. In addition, the nearly 5 W of "eye-safe", uncompressed output power around 1.96 μm may also be directly used for applications such as atmospheric remote sensing (e.g., of $CO_2$) or coherent Doppler lidar [23,24].

We thank T. Fortier, I. Hartl, K. Knabe, N. Newbury, J. Ye, J. Squier, and C. Durfee for valuable discussions and helpful contributions. This project was funded by the National Institute of Standards and Technology (NIST). F. A. acknowledges support from the NIST-ARRA fellowship program. As a contribution of NIST, an agency of the United States government, this work is not subject to copyright in the US. The mentioning of company and product names is for technical communication only and constitutes no endorsement by NIST.

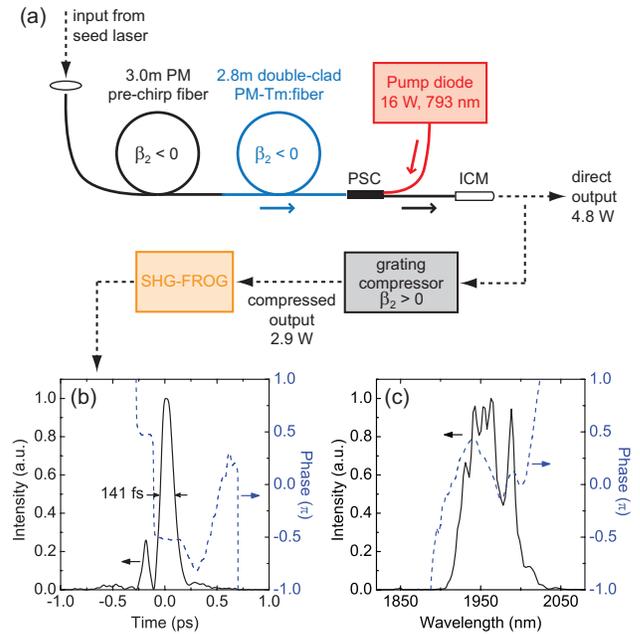

Fig. 2. (Color online) (a) Schematic of the polarization-maintaining double-clad Tm:fiber amplifier; PSC, pump-signal combiner; ICM, isolator/collimator module; solid lines indicate fiber path, dashed lines indicate free-space path. (b) Temporal intensity (black solid line) and phase (blue dashed line) of the compressed amplified pulse retrieved from the FROG measurement. (c) Corresponding spectral intensity (black solid line) and phase (blue dashed line).

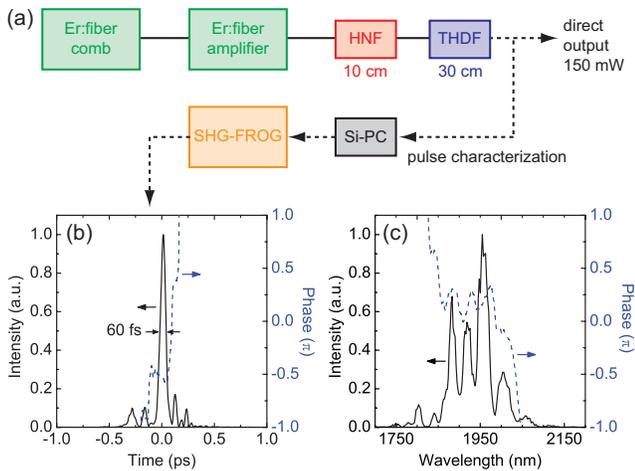

Fig. 1. (Color online) (a) Schematic of the setup generating the hybrid 2-µm frequency comb seed laser and the setup for pulse characterization consisting of silicon prism compressor (Si-PC) and SHG-FROG. HNF, highly nonlinear fiber; THDF, single-clad Tm/Ho-doped fiber; solid lines indicate fiber path, dashed lines indicate free-space path. (b) Temporal intensity (black solid line) and phase (blue dashed line) of the compressed seed pulse retrieved from the FROG measurement. (c) Corresponding spectral intensity (black solid line) and phase (blue dashed line).

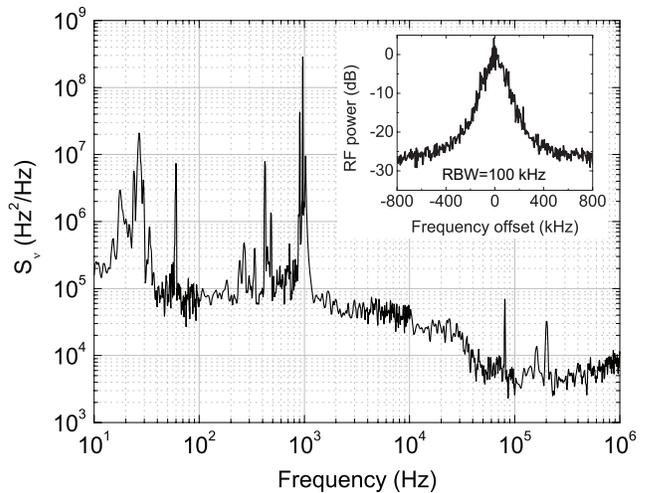

Fig. 3. Frequency noise spectral density $S_\nu(f)$ of the beat note (see inset) between the frequency-doubled hybrid comb after the double-clad PM-Tm:fiber amplifier and a frequency-stabilized cw laser at 976 nm.